\documentclass[12pt,subeqn,a4paper]{article}

\usepackage{amssymb,amsmath,amsfonts,amsthm, amscd, mathrsfs,helvet}
\usepackage{bm}
\usepackage{graphicx,verbatim}
\usepackage[all]{xy}
\usepackage{xcolor}
\usepackage{braket}
\usepackage{enumitem}

\usepackage{appendix}
\numberwithin{equation}{section}
\numberwithin{figure}{section}

\def\r2{\sqrt{2}}

\renewcommand{\baselinestretch}{1.4}

\newcommand{\half}{\frac{1}{2}}

\newcommand{\be}{\begin{equation}}
\newcommand{\ee}{\end{equation}}  
\newcommand{\pr}{\partial}

\newcommand{\Sym}{\text{Sym}}

\newcommand{\CP}{\mathbb{CP}}
\newcommand{\RP}{\mathbb{RP}}

\begin{document}
%zzz
\newcommand{\nd}[1]{/\hspace{-0.5em} #1}
\begin{titlepage}
%
%
%\begin{flushright}
%{May 2023} \\ 
%DAMTP-\\
%SWAT-\\ 
%hep-th/yymmnnn \\
%\end{flushright}
%
\begin{centering}

\vspace{.2in}

{\huge {\bf Neumann Boundary Condition for Abelian Vortices}}

\vspace{.4in}

{\large {\bf N.S. Manton and Boan Zhao}}\\\vspace{.2 in}
DAMTP, Centre for Mathematical Sciences \\ 
University of Cambridge, Wilberforce Road \\ 
Cambridge CB3 0WA, UK \\

\vspace{.8in}

{\bf Abstract} \\

\end{centering}

We study abelian BPS vortices on a surface $S$ with boundary,
which satisfy the Neumann boundary condition on the norm of the scalar field,
or equivalently, that the current along the boundary vanishes. These
vortices have quantised magnetic flux and quantised energy. Existence of
such vortices is manifest when $S$ is the quotient by a reflection of a
smooth surface without boundary, for example a hemisphere. The
$N$-vortex moduli space then admits an interesting stratification,
depending on the number of vortices in the interior of $S$ and the
number of half-vortices on the boundary. 

\end{titlepage}

\setcounter{tocdepth}{2}

\renewcommand{\baselinestretch}{1.1}\normalsize
\tableofcontents
\renewcommand{\baselinestretch}{1.4}\normalsize

\vspace{2em}
\section{Abelian BPS vortices and Neumann boundary condition}

Let $S$ be an oriented Riemannian surface with boundary $\partial
S$. We simplify by assuming that $\partial S$ is a single closed loop,
but the generalisation to a finite set of disjoint loops is
straightforward. We choose isothermal coordinates ${\bf x} =
(x^1,x^2)$ on $S$ and express the metric as
\be
ds^2 = \Omega(x^1,x^2) ((dx^1)^2 + (dx^2)^2) \,.
\ee
The area element is $\Omega \, d^2x$.

The simplest abelian gauge theory on $S$ with time-independent fields has a
single unit-charge scalar (Higgs) field $\phi$ and a gauge potential
$(a_1, a_2)$. BPS vortices are solutions of the Bogomolny equations
\begin{align}\label{BPS}
&D_1\phi + iD_2\phi = 0 \,, \nonumber\\
&F = \frac{\Omega}{2}(1 - \bar{\phi} \phi) \,,
\end{align}
where the gauge covariant derivative is
$D_i\phi = \partial_i\phi -ia_i\phi$ and the gauge field strength is
$F = \partial_1 a_2 - \partial_2 a_1$ \cite{Bog, book}. The magnetic
field (the coordinate-independent Hodge-dual of the field strength) is $B =
\frac{1}{\Omega} F$. By differentiating the second Bogomolny equation,
and using the first, one finds that the (electric) current, the source
for the magnetic field, is
\begin{equation}
J_i = -\frac{i}{2}(\bar{\phi}D_i \phi - \phi \overline{D_i\phi}) \,.
\end{equation}

Let us write
\begin{equation}
\phi = e^{\half h + i \chi} \,,
\label{phidecomp}
\end{equation}
where $h$ is gauge-invariant and single-valued, but $\chi$ is
gauge-dependent and has net winding $2\pi$ around each basic vortex
in the interior of $S$ (and winding $2\pi \mu$ around a vortex of
multiplicity $\mu$). Then $|\phi|^2 = e^h$ and $h$ has logarithmic
singularities at the vortex locations, where $\phi$ vanishes. The
current also simplifies to
\be
J_i = (\pr_i \chi - a_i)e^h \,.
\label{hcurrent}
\ee
Until section 3 we assume that no vortices are located on
the boundary.

For BPS vortices, there is a simple relationship between the current
$J_i$ and the contours of $h$. Let $D_1 \phi$ and $D_2 \phi$ denote the
components of the covariant derivative of $\phi$ tangent and normal
to a contour (for suitably chosen local, isothermal coordinates). The
first Bogomolny equation simplifies, using the decomposition
(\ref{phidecomp}), to
\be
\pr_1 \chi - a_1 = -\half \pr_2 h \,, \quad \pr_2 \chi - a_2 = 0 \,,
\label{TNderivs}
\ee
the right-hand side of the second equation being zero because
$\pr_1 h = 0$ along a contour of $h$. Using (\ref{hcurrent}) we see
that the normal component $J_2$ of the current vanishes, and the
tangential component is
\be
J_1 = -\half \pr_2 h \, e^h = -\half \pr_2 (e^h) = -\half \pr_2 |\phi^2| \,.
\ee
The current is therefore along the contours of $h$ and is non-zero, except
where $h$ is stationary in all directions.

It is straightforward to solve the first
Bogomolny equation using the decomposition (\ref{phidecomp}) to find
the gauge potential components, and deduce that
$F = -\half \nabla^2 h = -\half(\pr_1^2 + \pr_2^2)h$, away
from the singularities of $h$. The second Bogomolny equation then
reduces to the Taubes equation \cite{Tau} 
\be
\nabla^2 h +\Omega (1 - e^h) =
4\pi \sum_k \mu_k \, \delta^2({\bf x} - {\bf X}_k) \,,
\ee
where the delta functions arise from the logarithmic singularities and
$\mu_k$ is the multiplicity of the zero of $\phi$ at ${\bf X}_k$ -- a
positive integer. The total vortex number is $N = \sum_k \mu_k$.

Vortices on a surface $S$ with boundary have been studied before
in the context of superconductivity. Let $(t,n)$ be local isothermal
coordinates in the neighbourhood of the boundary $\pr S$, with $t$ a
coordinate along $\pr S$ anticlockwise, $n=0$ on $\pr S$, and $n$
positive in the interior. The usual boundary condition for
superconductors is the vanishing of the normal current, $J_n = 0$, so
the current remains inside the superconductor \cite{dGe}. It is easily
verified that this requires $D_n\phi$ to be a real multiple of $\phi$
at any point on $\pr S$ where $\phi \ne 0$. The stronger condition
$D_n\phi = 0$ is sometimes imposed. Vortices in superconductors are not
generally BPS -- they do not satisfy the Bogomolny equations -- so of
more relevance here is Nasir's study \cite{Nas} of BPS vortices
satisfying the Dirichlet boundary condition $|\phi| = 1$, which
included a numerical construction of $N=1$ and $N=2$ solutions on a
disc and on a square, with the vortices at various interior locations.
For these vortices, the energy is quantised, but the magnetic flux is not.

In this paper, we study BPS vortices satisfying the alternative Neumann
boundary condition, the vanishing of the tangential current on $\pr S$,
\begin{equation}
J_t = 0 \,,
\end{equation}
which requires $D_t\phi$ to be a real multiple of $\phi$. 
The vanishing of $J_t$ has consequences analogous to
what we found above for the current along contours of $h$. The first
Bogomolny equation implies that $D_t\phi + iD_n\phi = 0$, and using
the decomposition (\ref{phidecomp}) this simplifies to 
\be
\pr_t \chi - a_t = -\half \pr_n h \,,
\quad \pr_n \chi - a_n = \half \pr_t h \,.
\label{tnderivs}
\ee
As $J_t = (\pr_t \chi - a_t) e^h$ and $e^h$ is
everywhere non-zero on $\pr S$, the boundary condition $J_t = 0$ is equivalent
to $\pr_n h = 0$ on $\pr S$ (implying that $\pr_n |\phi|^2 = 0$). This
is the usual form of the Neumann boundary condition.

\section{Basic consequences of the Neumann condition}

With the Neumann boundary condition, the current is normal to the boundary,
but at the same time is along contours of $h$, so the contours of $h$ must
intersect the boundary orthogonally, except where $h$ is stationary
in both the tangential and normal directions. The normal current strength
is $J_n = (\pr_n \chi - a_n)e^h = \half \pr_t h \, e^h = \half
\pr_t(e^h)$, and is generally non-zero. However, the integrated current
crossing the boundary is
\be
\int_{\pr S} J_n \, dt = \half \int_{\pr S} \pr_t (e^h) \, dt = 0 \,.
\ee
It vanishes because $h$ is single-valued and has no
singularities on the boundary.

Now consider the magnetic flux. From (\ref{tnderivs}),
\be
\pr_t \chi - a_t = 0
\label{phasebound}
\ee
because of the Neumann boundary condition. By Stokes' theorem,
the total magnetic flux is
\be
\int_S B \, \Omega \, d^2x = \int_S F \, d^2x = \int_{\pr S} a_t \, dt \,, 
\label{flux}
\ee
so (\ref{phasebound}) implies that the flux is the increase of
$\chi$ around $\pr S$. Along a small loop $\gamma_k$ around the
zero of $\phi$ at ${\bf X}_k$, $\chi$ increases by $2\pi \mu_k$, so $\chi$
increases by $2\pi \sum_k \mu_k = 2\pi N$ around $\pr S$, where $N$
is the total vortex number (see Fig. 2.1). The Neumann boundary
condition, unlike the Dirichlet boundary condition,
therefore gives the familiar result that the magnetic flux for
an $N$-vortex solution of the Bogomolny equations is quantised and
equal to $2\pi N$. Having established flux quantisation, we can integrate
the second Bogomolny equation, and show in the standard way
that a necessary condition for an $N$-vortex solution to exist is
that the area $A$ of $S$ satisfies the Bradlow inequality
$A > 4\pi N$ \cite{Bra}.

\begin{figure}\label{contour_1}
\centering
\includegraphics[height=200pt]{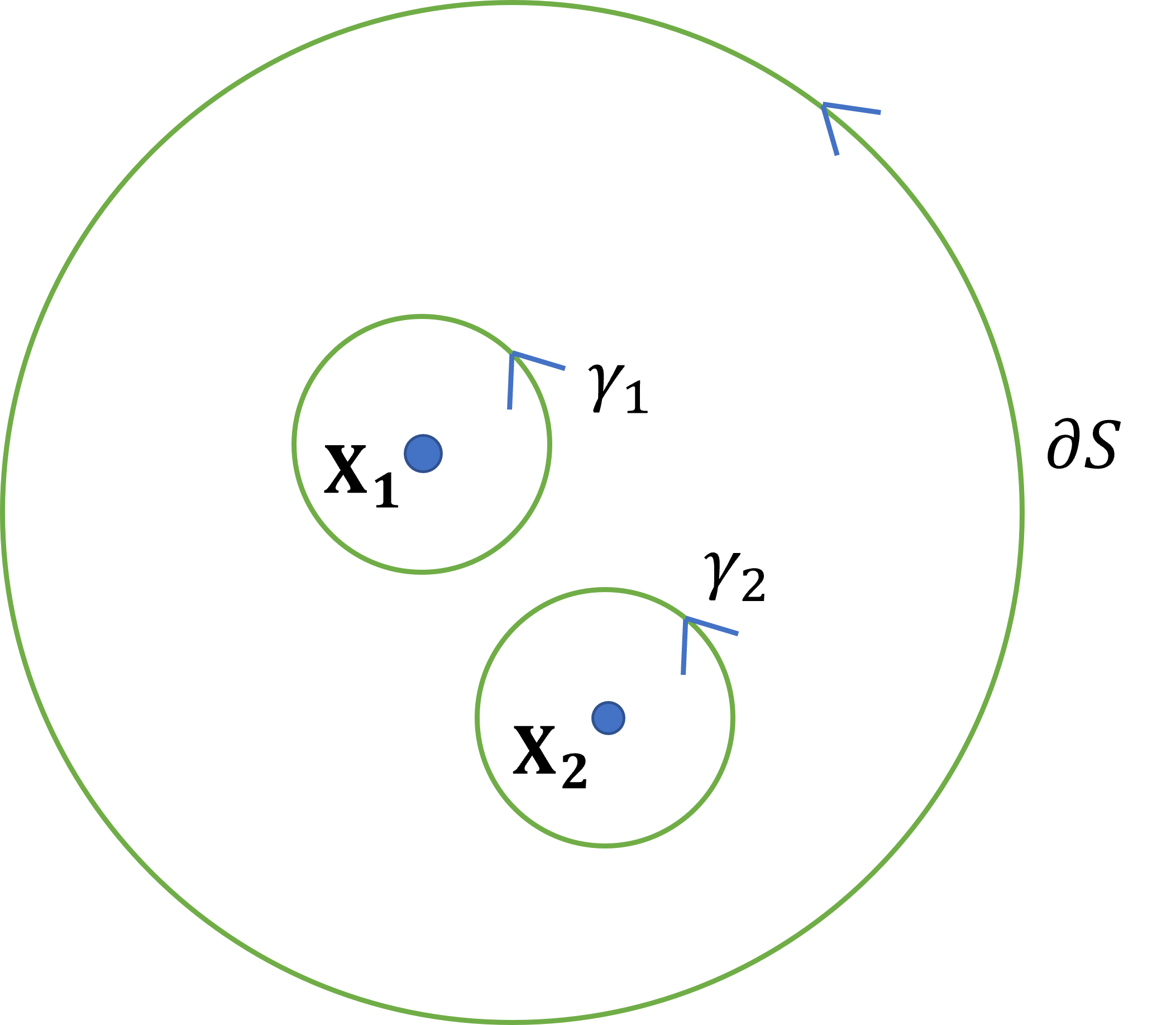}
\caption{Contours along the boundary $\pr S$ and around interior vortices.}
\end{figure}

Next, we show that the field energy
\begin{equation} \label{energy_functional}
E[\phi, a_i] = \int_S \half\left(\Omega^{-1}F^2
+ \overline{D_i\phi}D_i\phi + \frac{\Omega}{4}(1 - \bar{\phi}\phi)^2\right)
d^2x 
\end{equation}
is also quantised. By the standard Bogomolny rearrangement \cite{book},
\begin{align}\label{complete_the_square}
E =& \int_S \half\left(\Omega^{-1}\left(F -
\frac{\Omega}{2}(1- \bar{\phi}\phi)\right)^2 +
(\overline{D_1\phi} - i\overline{D_2\phi})(D_1\phi + iD_2\phi)
+ F \right) d^2x \nonumber\\
&+\int_S \half\left(i\overline{D_2\phi}D_1\phi
- i\overline{D_1\phi}D_2\phi -F\bar{\phi}\phi\right) \, d^2x \,. 
\end{align}
If the Bogomolny equations (\ref{BPS}) are satisfied, the integral
in the first line reduces to $\pi N$, half the integral of $F$. Then, since
\begin{equation}
-F\bar{\phi}\phi = i\bar{\phi}D_2D_1\phi - i\bar{\phi}D_1D_2\phi
\end{equation}
the second line of the energy can be written as
\begin{equation}
\int_S \half\left(i\partial_2(\bar{\phi}D_1\phi)
- i\partial_1(\bar{\phi}D_2\phi)\right) \, d^2x \,.
\label{Eboundary}
\end{equation}
For vortices on a plane, or on a compact surface without boundary,
this integrand has a convenient form, even though it is
not manifestly real. However, here we need a variant. We note that
\be
\Re(i\bar{\phi}D_1\phi))
= \frac{i}{2}(\bar{\phi}D_1\phi - \phi\overline{D_1\phi}) = -J_1
\ee
and
\be
\Im(i\bar{\phi}D_1\phi)) = \half(\bar{\phi}D_1\phi + \phi\overline{D_1\phi})
= \half \pr_1(|\phi|^2) \,,
\ee
and similarly for $i\bar{\phi}D_2\phi$. The imaginary
part of the integrand of (\ref{Eboundary}) is therefore \linebreak
$\frac{1}{4}\left(\pr_2\pr_1(|\phi|^2) - \pr_1\pr_2(|\phi|^2)\right)$,
which vanishes, so the integrand is actually real
and equal to $\half(\pr_1 J_2 - \pr_2 J_1)$. By Stokes' theorem,
(\ref{Eboundary}) is therefore half the integral of $J_t$ around $\pr S$,
and this vanishes because of the Neumann condition. In
conclusion, a BPS $N$-vortex solution satisfying the Neumann boundary
condition has quantised energy $\pi N$.

We can prove the uniqueness of BPS vortex solutions with prescribed
zeros of $\phi$, satisfying the Neumann boundary condition, but
this is deferred to section 6. However, we do not offer a general proof of
vortex existence.

\section{Quotient surfaces arising from a reflection}

We now specialise to a large class of surfaces $S$ on which the
existence of BPS vortices satisfying the Neumann boundary condition is
clear. Let $\tilde S$ be a smooth closed, oriented
Riemannian surface without boundary that admits a reflection
symmetry $\sigma$. Let $E$ be the fixed-point set of $\sigma$. The
quotient of $\tilde S$ by $\sigma$ is a surface $S$ with boundary
$\pr S = E$. Note that as $\sigma$ is an isometry, $E$ is a geodesic
on $\tilde S$. $E$ is also a geodesic on $S$, provided we
restrict variations of $E$ to be into the interior of $S$.
In this section we will allow vortex locations to be either in
the interior of $S$ or on its boundary $E$.

There are plentiful examples. The simplest $\tilde S$ is a sphere, with
$\sigma$ reflecting the northern to the southern hemisphere. Then $S$ is a
hemisphere with boundary $E$ the equator. Alternatively, $\tilde S$
could be a kidney-shaped surface. This has two independent reflection
symmetries, and each quotient is diffeomorphic to a hemisphere, with $E$
a single closed loop. For a doughnut sliced in half, $E$ is the union of two
circles. Closed surfaces of revolution, e.g. pear-shaped surfaces, sliced in
half by a plane containing the rotation axis, give further examples.

Not all surfaces are of this type. Even though a surface with boundary
can always be glued to a reflected copy of itself, the result is not
generally smooth. Recall that for a smooth, oriented surface
$\tilde S$ of genus $g$, the Gauss curvature $K$ has integral
\be
\frac{1}{2\pi} \int_{\tilde S} K \, \Omega \, d^2x = 2 - 2g \,.
\ee
If $S$ is the quotient of such a surface by a reflection, then
the corresponding integral over $S$ is $1 - g$, i.e. an integer no
greater than 1. For $g=0$, the integral is 1, so $S$ can
be a round hemisphere, but not a planar disc.

The existence of BPS vortices on a smooth, closed surface $\tilde S$ is well
established \cite{Bra,Gar}. If $\tilde S$ has area $\tilde A$, then
$\tilde N$-vortex solutions exist for all $\tilde N$ such that
$4\pi \tilde N < \tilde A$, and have total flux $2\pi \tilde N$.
Moreover the zeros of the scalar field $\phi$ can be at arbitrary locations,
with multiplicities summing to $\tilde N$. For
given locations and multiplicities, the solution is unique up to gauge
equivalence. Generically there are $\tilde N$ unordered, simple zeros,
so the moduli space of solutions is $\Sym^{\tilde N}(\tilde S)$, the
$\tilde N$th symmetrised power of $\tilde S$. Note that an oriented Riemannian
surface is a (1-dimensional) complex manifold, and its symmetrised
$\tilde N$th power is also complex. In particular, for
$\tilde N$-vortices on a surface diffeomorphic to ${\CP}^1$ (a sphere),
the moduli space is diffeomorphic to ${\CP}^{\tilde N}$.

Vortex solutions on a surface $S$ of the quotient type, satisfying
the Neumann boundary condition on $E$, are obtained simply from all the
vortex solutions on $\tilde S$ that respect the reflection
symmetry $\sigma$. The vortex locations on $\tilde S$ need
to be a combination of distinct pairs mapped into each other
by $\sigma$, with equal multiplicities, and unpaired
locations on $E$, with arbitrary multiplicities. By the uniqueness of
vortex solutions for given locations and multiplicities, the action of
$\sigma$ then gives the same solution (up to gauge equivalence).
So the field configuration is invariant under $\sigma$,
and therefore well-defined on $S$.

\begin{figure}\label{contour_2}
\centering
\includegraphics[height=200pt]{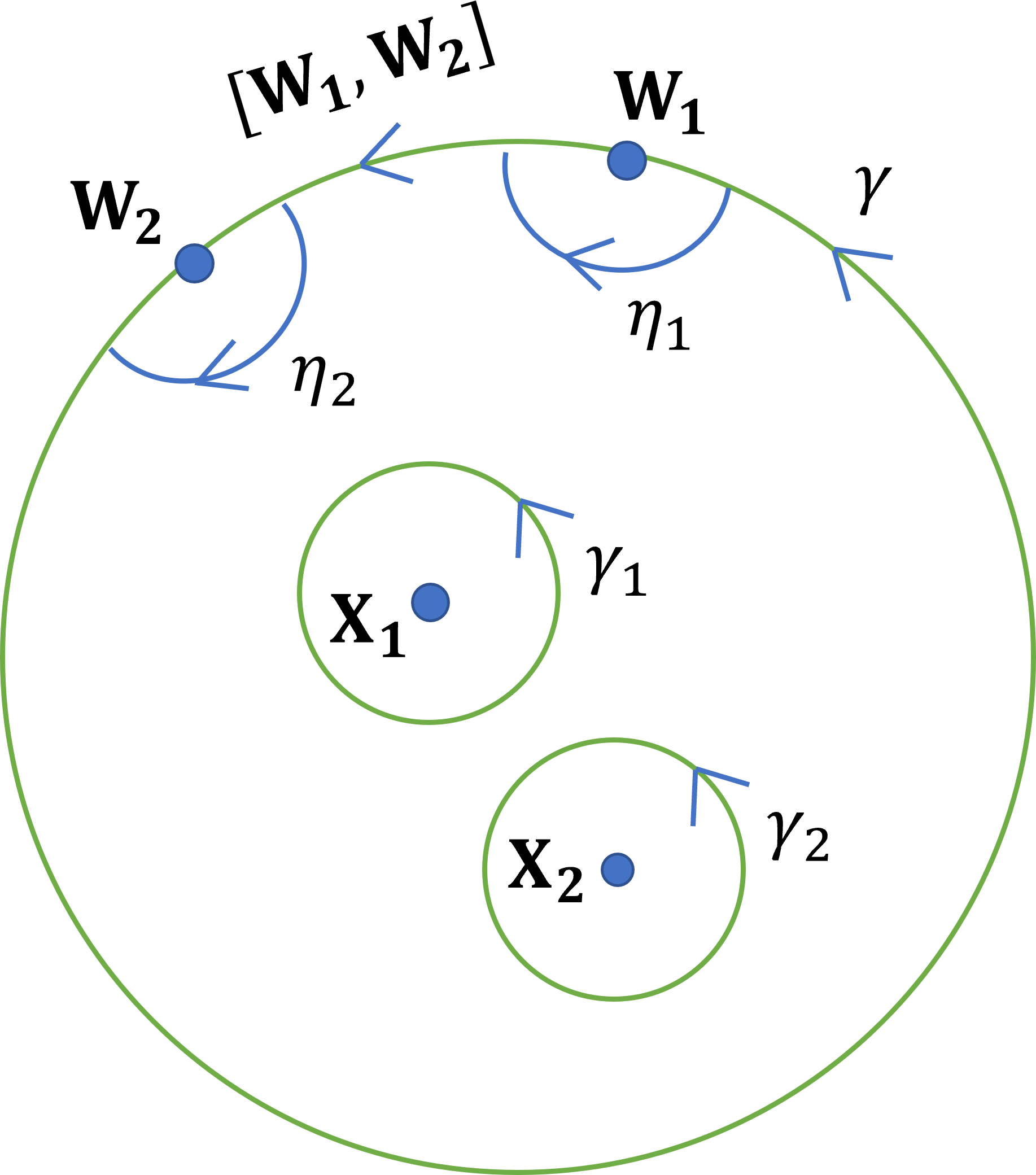}
\caption{Choice of the contours $\gamma,\gamma_k,\eta_j$. $\gamma_k$
is a small circle surrounding the interior zero ${\bf X}_k$. $\eta_j$ is
a small semicircle surrounding the boundary zero ${\bf W}_j$, and
$[{\bf W}_j,{\bf W}_{j+1}]$ denotes the truncated segment of the
boundary between ${\bf W}_j$ and ${\bf W}_{j+1}$, so $\gamma =
\sum_j ([{\bf W}_j , {\bf W}_{j+1}] + \eta_j)$.}
\end{figure}

In particular, on $\tilde S$ the gauge-invariant field $h$ is strictly
invariant under the reflection, so descends to $S$ and obeys
the Taubes equation. Moreover, as $h$ has the same value at pairs of
points related by $\sigma$ close to $E$, the normal derivative of $h$
on $E$ must vanish. The tangential current $J_t$ on $E$ also vanishes;
heuristically this is because the vortices in the two halves of $\tilde S$
generate currents in opposite directions along $E$, so the net current
is zero. This quotient construction therefore gives BPS vortex solutions
on $S$, satisfying the Neumann boundary condition. The solutions have
ordinary vortices in the interior of $S$, with integer multiplicity,
but additionally can have half-vortices on the boundary. The total
vortex number on $S$ is of the form $N = \half \tilde N$, with
$\tilde N$ an integer, so $N$ can be a half-integer. The magnetic flux
through $S$ is $2\pi N$. For such vortices to exist when $S$ has area
$A$, it is necessary that $4\pi N < A$.  

Let us clarify what is meant by a half-vortex. Roughly speaking, it is
a reflection-symmetric vortex located on $E \subset \tilde S$, sliced
in half. The total flux on $\tilde S$ due to a unit-multiplicity vortex on
$E$ is $2\pi$. On $S$ the half-vortex has flux $\pi$. Similarly, the
energy of the half-vortex is $\half \pi$, half of the energy on $\tilde S$.
Half-vortices on $E$ can also occur with higher multiplicity.

We can more directly calculate the total flux of an $N$-vortex
solution on $S$, using the boundary condition and Stokes' theorem.
We simplify, by assuming again that the boundary of $S$ is a single
closed loop. As before, the flux is the integral of $a_t$ along the
boundary. Let ${\bf X}_k$ be the interior zeros of $\phi$, with total
multiplicity $q$, and let ${\bf W}_j$ be the boundary zeros. The
total multiplicity of boundary half-vortices is $2(N-q)$. Choose
a contour $\gamma$ that coincides with the boundary $E$ except for small
semicircles $\eta_j$ around the boundary zeros, and choose small
circular contours $\gamma_k$ around the interior zeros. See Fig. 3.1
for more details.

Now
\begin{equation}
\int_\gamma \pr_t\chi \, dt = \sum_j \int_{[{\bf W}_j,{\bf W}_{j+1}]}
\pr_t \chi \, dt + \sum_j \int_{\eta_j} \pr_t \chi \, dt \,,
\label{phaseboundary}
\end{equation}
where $[{\bf W}_j, {\bf W}_{j+1}]$ denotes the truncated segment of
$E$ between ${\bf W}_j$ and ${\bf W}_{j+1}$, excluding the part
inside the semicircles $\eta_j$ and $\eta_{j+1}$.
Clearly, the left-hand side is the sum of the integrals around the
contours $\gamma_k$, and equals $2\pi q$. In the first term
on the right-hand side, on the segments $[{\bf W}_j, {\bf W}_{j+1}]$
we can replace $\pr_t \chi$ by $a_t$ because of the Neumann boundary
condition. The second term is $-2\pi(N-q)$ because the contours $\eta_j$ run
clockwise around each boundary zero, and each half-vortex contributes
$-\pi$, by reflection symmetry. 

We now take the limit as the semicircles $\eta_j$ shrink to zero size,
and the segments $[{\bf W}_j, {\bf W}_{j+1}]$ stretch fully between
${\bf W}_j$ and ${\bf W}_{j+1}$. $a_t$ approaches zero as each
boundary zero is approached along $E$, so the limit is
straightforward. Then (\ref{phaseboundary}) reduces to
\be
2\pi q = \sum_j \int_{[{\bf W}_j,{\bf W}_{j+1}]} a_t \, dt - 2\pi(N-q) \,.
\label{fluxboundary}
\ee
The sum over the boundary segments becomes the whole boundary $E$, so
the sum of integrals of $a_t$ is the total magnetic flux. $2\pi q$
cancels, so the total flux is $2\pi N$, as before. 

\section{The moduli space and its stratification}

The moduli space of $\tilde N$-vortex solutions on $\tilde S$, assuming its
area is sufficient for such solutions to exist, is a complex manifold
of complex dimension $\tilde N$. The reflection $\sigma$ is orientation
reversing on $S$, i.e. antiholomorphic, so the subspace of
reflection-symmetric vortex solutions, which we have identified with
the vortices on $S$ satisfying the Neumann boundary condition on the
boundary $E$, is a real subspace of real dimension $\tilde N = 2N$.

We can account for this as follows. Consider the generic solutions where
the multiplicity of each vortex on $\tilde S$ is 1. Assume that on $S$
there are $q$ interior vortices and $2(N - q)$ boundary vortices. An
interior vortex has two real degrees of freedom as its location is
arbitrary. A boundary half-vortex has one real degree of freedom as it
is constrained to move along the boundary $E$. The total number of
degrees of freedom is $2q + 2(N-q) = 2N$, independent of $q$.

For the remainder of this section, we assume that $S$ is a hemisphere,
or conformal to a hemisphere. We choose coordinates $(x^1,x^2)$ and introduce
the complex coordinate $z = x^1 + i x^2$, so that the equator $E$ is the real
$x^1$-axis, and the hemisphere is covered by the upper half-plane
$x^2 \ge 0$ extended to include $\infty$. The sphere $\tilde S$ is
covered by the entire, extended complex plane. Now an $\tilde N$-vortex
solution on $\tilde S$ is uniquely specified by the polynomial $P(z)$
whose zeros occur at the vortex locations \cite{Tau}; the multiplicity
of a zero determines the multiplicity of the vortex. $P(z)$ cannot
be identically zero and an overall constant factor has no effect on the
zeros. Generically, $P$ has degree $\tilde N$, but if the degree is
less than this, it is understood that the deficit in degree is the
vortex multiplicity at $\infty$.

\begin{figure}\label{half_vortex}
\centering
\includegraphics[height=200pt]{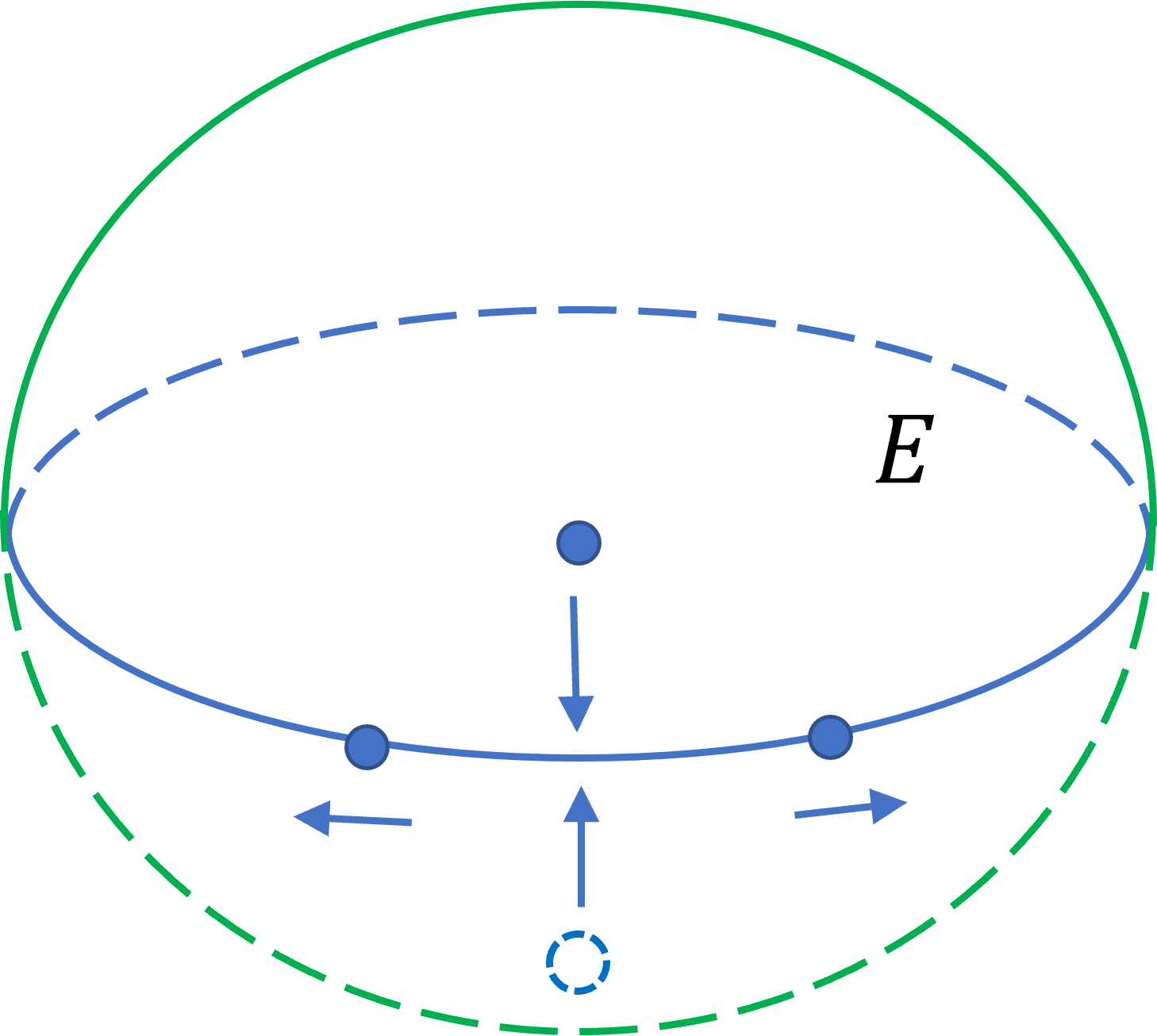}
\caption{Vortex approaching boundary $E$ and splitting into two half-vortices.}
\end{figure}

Vortices on $S$ are obtained by imposing the reflection symmetry $\sigma$,
which is equivalent to $z \leftrightarrow \bar z$. The polynomials $P(z)$
satisfying this, after multiplication by an overall phase factor if
necessary, are those with real coefficients. The set of zeros
is then a union of complex conjugate pairs, and zeros on the
real axis. This is what we expect. The vortices on $\tilde
S$ are pairs of vortices not on $E$, related by the reflection, together with
vortices on $E$. On $S$ there are vortices at arbitrary
locations in the interior (the roots of $P$ in the upper half-plane)
and half-vortices on $E$ (the real roots of $P$ together with
the roots at $\infty$).

${\cal M}_N$, the moduli space of $N$-vortices on $S$ (with $N$ an
integer or half-integer) is therefore the
real projective space ${\RP}^{2N}$, the space of $2N+1$ real coefficients
of a polynomial of degree $2N$, modulo multiplication by a common
real constant. This moduli space has a natural
stratification. Let $q$ be an integer in the range $0 \le q \le [N]$,
where $[N]$ is the integer part of $N$. The $q$th stratum
${\cal M}_{N,q}$ is the space of vortices where there are $q$
vortices in the interior of $S$ and $2(N - q)$ half-vortices on the
boundary. Mathematically, ${\cal M}_{N,q}$ is the projective
space of real polynomials of degree $2N$ with $q$ pairs of
complex roots and the remaining roots real. ${\cal M}_N$ is the
disjoint union of the strata ${\cal M}_{N,q}$.
The strata join continuously, as ${\RP}^{2N}$ is connected, and all the
strata have the same dimension. Almost everywhere, the join occurs when an
interior unit-multiplicity vortex moves to the boundary, and then
splits into two half-vortices that move away from each other along
the boundary. This is a smooth process. Viewed from $\tilde S$, the
first vortex is colliding with its reflection in $E$ (see Fig. 4.1). As is
familiar for vortices on a smooth compact surface, in such a collision the
vortices scatter at right angles if their local centre of mass is
at rest \cite{Sam}. In the reverse process on $S$, two half-vortices
collide on $E$, and a vortex moves into the interior. In the
absence of such a collision, a half-vortex is constrained to move
along $E$. 

\section{Examples}

\begin{figure}\label{example_2}
\centering
\includegraphics[height=250pt]{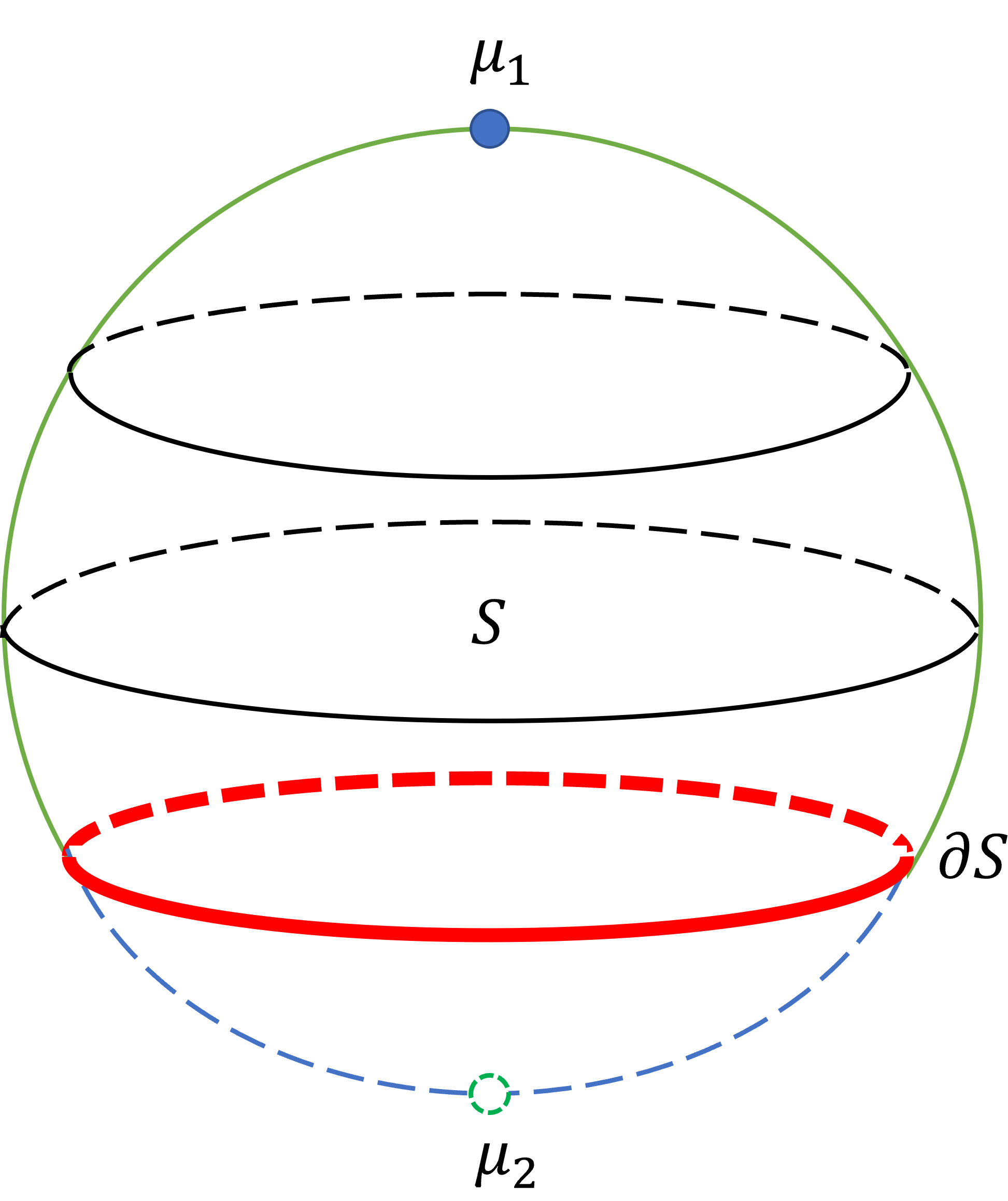}
\caption{Vortex of multiplicity $\mu_1$ at the north pole of a sphere
with a cap removed. The boundary $\pr S$ is shown bold. On the
complete sphere there is also a vortex of multiplicity $\mu_2 < \mu_1$
at the south pole. The contours of $h$ are parallel circles of latitude.}
\end{figure}

Here, we give two examples of vortices satisfying the Neumann boundary
condition. The first example (Fig. 5.1) is a vortex located at the north pole
of a round sphere with a cap around the south pole removed. The resulting
surface $S$ is not generally the quotient by a reflection of any
smooth surface. Consider a BPS vortex configuration on the complete
sphere with a vortex of multiplicity $\mu_1$ at the north pole and a
vortex of multiplicity $\mu_2 < \mu_1$ at the south pole. The sphere has to
have area greater than $4\pi(\mu_1 + \mu_2)$. $h$ is
rotationally-invariant and has (negative) logarithmic singularities at the
poles. $h$ attains its maximum on some circle below the equator. On
this circle, $h$ is stationary and the Neumann boundary condition is
satisfied. So we take this circle to be the boundary $\partial S$.
The vortex number of this solution on $S$ is $N=\mu_1$. Note
that in this example the contours of $h$ do not intersect the boundary
orthogonally, because $h$ is stationary in all directions on the boundary. 

\begin{figure}
\centering
\includegraphics[height=200pt]{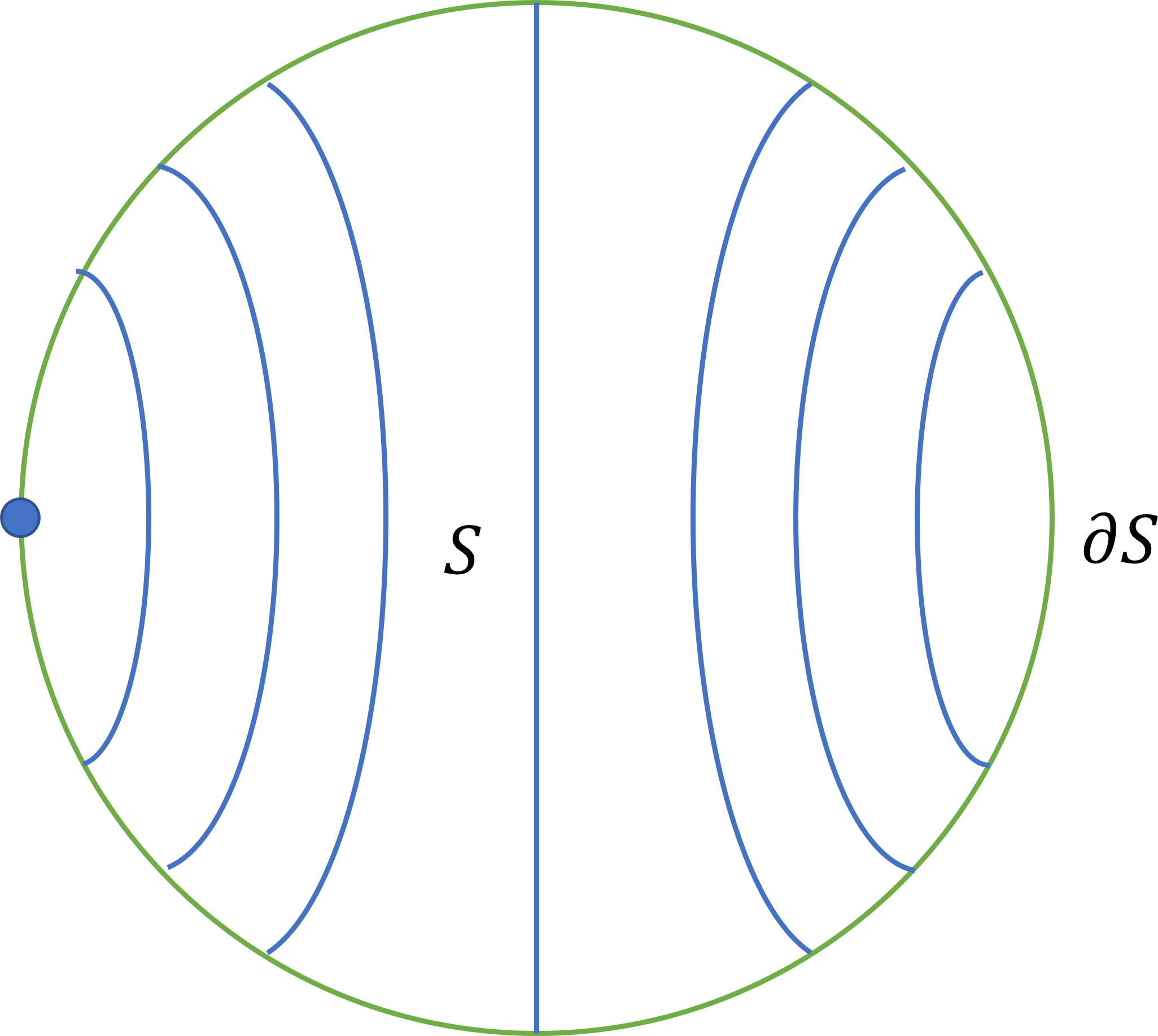}
\caption{Half-vortex on the boundary of a hemisphere.
The semi-circular contours of $h$ intersect the boundary orthogonally.}
\end{figure}

The second example (Fig. 5.2) is a half-vortex on the boundary of a
round hemisphere with area greater than $2\pi$. The vortex number is
$N = \half$. The half-vortex can be regarded as a unit-multiplicity
vortex on the equator of a complete sphere, sliced in half. So the
contours of $h$ are semi-circles of latitude relative to the pole at
the vortex location. These contours intersect the boundary orthogonally.

\section{Uniqueness of BPS vortex solutions}

Here we consider BPS vortices on a general surface $S$ with boundary
$\pr S$ (not necessarily the quotient of $\tilde S$ by a
reflection) satisfying the Neumann boundary condition, and prove the
uniqueness of a solution with prescribed zeros. The proof essentially
follows \cite{Nas}. Again, let $z = x^1 + ix^2$. We use the
Taubes equation in the form
\begin{align}\label{taubes}
\nabla^2 h &= \Omega(e^h - 1) + 4\pi \sum_k \delta^2 (z - z_k)\nonumber\\
\partial_nh|_{\pr S} &= 2\pi\sum_j \delta^1 (z - w_j) \,,
\end{align}
where $z_k$ are the interior zeros of $\phi$ and $w_j$ are the
boundary zeros (assumed here to be of unit multiplicity).
$\delta^2$ denotes a 2d Dirac delta in the interior of $S$ and
$\delta^1$ a 1d Dirac delta on the boundary. As before, $\pr_n$
denotes the inward normal derivative.

To verify the coefficient multiplying
the 1d Dirac delta, we choose $z$ to vanish at the boundary
point $w_j$, with the boundary locally the $x^1$-axis, and the interior
$x^2 >0$, so the metric is locally $(dx^1)^2+(dx^2)^2$ and the inward
normal derivative is $\pr_2$. Near $w_j$, $|\phi|\sim |z|$ up to a
multiplicative constant, so $h\sim 2\log|z|$ up to an additive
constant and hence
\begin{equation}
\partial_n h \sim 2\pr_2\log|z| =
\frac{2x^2}{(x^1)^2 + (x^2)^2} \,.
\end{equation}
This vanishes for $x^2 = 0$ when $x^1 \ne 0$, but integrating along a
short segment parallel to the boundary, with $x^2$ small, we find
\begin{equation}
\int \partial_nh \, dx^1 \sim
\int \frac{2x^2}{(x^1)^2 + (x^2)^2} \, dx^1 = 2\pi \,,
\end{equation}
as required.

We can remove the Dirac deltas from \eqref{taubes}
by shifting $h$ by a simple, singular function $H$ with appropriate
logarithmic singularities satisfying
\begin{align}
\nabla^2 H &= 4\pi\sum_k \delta^2 (z - z_k)\nonumber\\
\partial_nH|_{\pr S} &= 2\pi\sum_j \delta^1(z - w_j) + H_0\,,
\end{align}
where $H_0$ is a smooth function defined on the boundary $\partial S$.
Now we pick a smooth function on $S$ whose inward normal derivative on
the boundary equals $-H_0$. Let $G$ be the sum of this smooth function
and $H$. It satisfies
\begin{align}
\nabla^2 G &= 4\pi\sum_k \delta^2 (z - z_k) + G_0\nonumber\\
\partial_nG|_{\pr S} &= 2\pi\sum_j \delta^1(z - w_j) \,,
\end{align}
where $G_0$ is a smooth function in the interior of $S$, continuous
up to the boundary. One should be aware that it is impossible to
eliminate $G_0$ due to the divergence theorem
\begin{equation}
\int_S\nabla^2 G \, d^2x = -\int_{\pr S} \partial_nG \, dt \,.
\end{equation}
Although $G$ is singular, $e^G$ is a continuous function (up to
the boundary) and vanishes at the zeros of $\phi$.

Now define $f = h - G$; it satisfies
\begin{align}\label{taubes2}
\nabla^2 f &= \Omega \, e^{f + G} - \Omega - G_0\nonumber\\
\partial_n f|_{\pr S} &= 0 \,.
\end{align}
Solutions to this modified Taubes equation and Neumann boundary
condition for $f$ are precisely the critical points of the functional
\begin{equation}
I[f] = \int_S \left(\frac{1}{2}|\nabla f|^2
+ \Omega \, e^{f + G} - \Omega f - G_0 f\right) d^2x \,.
\end{equation}
This is because
\begin{align}
\frac{d}{d\lambda}I[f + \lambda \delta f]|_{\lambda = 0}
&= \int_S \left(\nabla f \cdot \nabla\delta f +
\Omega \, e^{f + G}\delta f -\Omega \delta f - G_0 \delta f \right)
\, d^2x\nonumber\\
&= \int_S \delta f \left(-\nabla^2 f + \Omega \, e^{f + G} - \Omega
- G_0 \right) \, d^2x \,.
\end{align}
The boundary term disappears due to the Neumann condition on $f$.

We will next show that every critical point of the functional $I$ is a local
minimum and that $I$ is convex. First we compute the second
derivative at a critical point,
\begin{equation}
\frac{d^2}{d\lambda^2}I[f + \lambda \delta f]|_{\lambda = 0}
= \int_S \left(|\nabla \delta f|^2 + \Omega \, e^{f + G}(\delta f)^2
  \right) \, d^2x \,.
\end{equation}
This is positive for $\delta f \ne 0$. The strict convexity of $I$, i.e.
\begin{equation}
I[\nu f_1 + (1-\nu)f_2] < \nu I[f_1] + (1-\nu)I[f_2]
\quad {\rm for} \ \nu \in (0,1), \, f_1\neq f_2 \,,
\end{equation}
follows from the strict convexity of the integrand of $I$ at every
point of $S$ (except the zeros of $\phi$). Equality holds
only when $f_1 = f_2$.

Now we are ready to prove the uniqueness of the solution to the
modified Taubes equation (\ref{taubes2}). Assume that $f_1,f_2$ are
two distinct solutions. Without loss of generality we can assume
$I[f_2]\geq I[f_1]$. The strict convexity of $I$ implies that
$I[\nu f_1 + (1-\nu)f_2] < I[f_2]$ for $\nu \in (0,1)$. However,
positivity of the second derivative of $I$ implies that
$I[\nu f_1 + (1-\nu)f_2] > I[f_2]$ when $\nu$ is sufficiently close to
0. This contradiction proves the uniqueness of $f$ and hence the
uniqueness of $e^h =|\phi|^2$ given the zeros of $\phi$.

The rest is simple. Given two vortex solutions
$(a_i^{(1)},\phi^{(1)}), (a_i^{(2)}, \phi^{(2)})$ with the same zeros and
same norm $|\phi^{(1)}|^2 = |\phi^{(2)}|^2$, we need to show that they
are gauge equivalent. Let $U: S - \{z_1,\dots;w_1,\dots\}\to
U(1)$ be the gauge transformation such that $U\phi^{(1)}= \phi^{(2)}$.
$U$ is defined away from the zeros and we need to show that it extends
continuously to the zeros. The first Bogomolny equation implies that
$\phi$ has the expansion around the interior zero $z_k$,
\be
\phi = C(z - z_k)^{\mu_k} + O(|z -z_k|^{\mu_k + 1})
\ee
for some nonzero $C$. As a result,
$\lim_{z\to z_k}\phi^{(1)}(z)/\phi^{(2)}(z)$ exists and so $U$ can be
continuously extended to $z_k$. A similar argument works for the
boundary zeros $w_j$. Hence $U$ is a well-defined gauge transformation
on $S$ that maps $\phi^{(1)}$ to $\phi^{(2)}$. The first
Bogomolny equation implies that $a_i$ is uniquely determined by
$\phi$ and its first partial derivatives, so $U$ gauge transforms
$a_i^{(1)}$ to $a_i^{(2)}$.

\section{Conclusions}
In this paper, we have studied the Neumann boundary condition for
abelian BPS vortices. The boundary condition has a number of nice
consequences such as the quantisation of magnetic flux, as well as of
energy. A novel feature of this boundary condition is the appearance
of half-vortices that are constrained to move on the boundary unless they
collide with other half-vortices. We have proved the existence of
solutions combining interior vortices and boundary half-vortices on
surfaces that are the quotient by a reflection of a smooth compact
surface without boundary, and have given some examples. Future
directions include the proof of existence (and numerical construction)
of vortex solutions on general surfaces with boundary. It is not clear if
the boundary fractional-vortices are necessarily half-vortices in this case.

Work on the generalisation of the Neumann boundary condition to abelian
multi-scalar vortices, and its consequences, is in progress.

\section*{Acknowledgements}

We thank Claude Warnick and Zexing Li for helpful discussions on the Taubes
equation. BZ is funded by a Trinity College, Cambridge internal
graduate studentship.

%\bibliographystyle{plain}
%\bibliography{VortNeumann.bib}

\vspace{.5cm}

\begin{thebibliography}{99}

\bibitem{Bog} E.~B. Bogomolny,
The stability of classical solutions, 
\textit{Sov. J. Nucl. Phys.} \textbf{24}, 449 (1976).

\bibitem{book} N. Manton and P. Sutcliffe,
\textit{Topological Solitons},
Cambridge, Cambridge University Press, 2004.

\bibitem{Tau} C.~H. Taubes, 
Arbitrary $N$-vortex solutions to the first order Ginzburg--Landau equations,
\textit{Commun. Math. Phys.} \textbf{72}, 277 (1980).

\bibitem{dGe} P.~G. de Gennes,
\textit{Superconductivity of Metals and Alloys},
New York, Benjamin, 1966.

\bibitem{Nas} S.~M. Nasir,
Study of Bogomol'nyi vortices on a disk,
\textit{Nonlinearity} \textbf{11}, 445 (1998).

\bibitem{Bra} S.~B. Bradlow, 
Vortices in holomorphic line bundles over closed K\"ahler manifolds,
\textit{Commun. Math. Phys.} \textbf{135}, 1 (1990).

\bibitem{Gar} O. Garc\'{\i}a-Prada, 
A direct existence proof for the vortex equations over a compact 
Riemann surface,
\textit{Bull. London Math. Soc.} \textbf{26}, 88 (1994).

\bibitem{Sam} T.~M. Samols, 
Vortex scattering,
\textit{Commun. Math. Phys.} \textbf{145}, 149 (1992).

\end{thebibliography}

\end{document}